\documentstyle[epsf,12pt]{article}

\newlength{\figsize}
\begin{document}

\begin{titlepage}

\begin{tabbing}
\` Oxford preprint OUTP-98-88P \\
\end{tabbing}
 
\vspace*{.1in}
 
\begin{center}
{\large\bf Glueball masses and other physical properties of\\
             SU(N) gauge theories in D=3+1:\\
\vspace*{.1in}
a review of lattice results for theorists\\}
\vspace*{.7in}
{Michael J. Teper\\
\vspace*{.3in}
\vspace{0.10cm}
Theoretical Physics, University of Oxford,\\
1 Keble Road, Oxford, OX1 3NP, U.K.\\
}
\end{center}

\vspace*{0.8in}

\begin{center}
{\bf Abstract}
\end{center}

We summarise what lattice simulations have to say about
the physical properties of continuum SU(N) gauge theories 
in 3+1 dimensions. The quantities covered are: 
the glueball mass spectrum, the confining string tension, 
the temperature at which the theory becomes deconfined, 
the topological susceptibility, the value of the scale 
$\Lambda_{\overline{MS}}$ that governs the rate at which the 
coupling runs and the $r_0$ parameter that characterises
the static quark potential at intermediate distances.

\end{titlepage}

\setcounter{page}{1}
\newpage
\pagestyle{plain}

\section{Introduction}
\label		{intro}

In recent years there has been a resurgence of interest
in developing analytic techniques for calculating
the physics of non-Abelian gauge theories.
Examples include approaches using light cone quantisation
\cite{dalley}
and the AdS calculations
\cite{AdS}
based on the conjecture by Maldacena
\cite{Malda}.
In practice all such calculations make approximations and 
so, in order to evaluate the accuracy of the approach being 
used, comparison with known results is very useful.
In this context `known results' means whatever has been learned
from the computer simulation of these theories. The purpose
of this review is to provide a readily accessible summary
for such theorists of what has been learned from lattice calculations 
about the continuum properties of non-Abelian gauge theories in 
3+1 dimensions.  

The quantities I shall focus upon are the glueball mass spectrum,
the confining string tension $\sigma$, 
the deconfining temperature $T_c$,
the $\Lambda_{\bar{MS}}$ parameter that sets the scale in
perturbative calculations, a parameter called $r_0$ that
is a characteristic scale of the static quark potential at
intermediate distances, and the topological susceptibility $\chi_t$.

There exists substantial information on these quantities
for SU(2) and SU(3) gauge theories. For many of these quantities
lattice calculations exist for a large enough range of
lattice spacings that a controlled extrapolation to the
continuum limit is possible. This continuum physics is 
summarised in Section 2. For some other quantities accurate 
calculations do exist but only for one or two values of the 
lattice spacing, $a$, and so a reliable continuum extrapolation
is not possible. Where such calculations have been performed 
for values of $a$ that are very small, the lattice
corrections to mass ratios should also be small. This applies
in particular to some of the heavier states of the mass spectrum,
and we list these in Section~\ref{su2su3_latt}. The reader should
be cautious in using the latter numbers: most of them 
`should' be close to their corresponding continuum values
but they might not be. 

While SU(2) and SU(3) gauge theories are of obvious
relevance, many analytic approaches
find it simplest to calculate the $N \to \infty$
limit of SU($N$) gauge theories; and this limit is, in any
case, of particular theoretical interest. So in the last 
section I will mention some very preliminary SU(4)
calculations and discuss what all this tells us about the
large-$N$ limit.

Although this review restricts itself to 3+1 dimensions, 
we note that from the point of view of 
most of these analytic approaches, non-Abelian gauge theories 
in 2+1 dimensions are equally interesting and
challenging. Here there is a recent lattice calculation of 
the glueball mass spectrum and string tension, in units of
the scale provided by the coupling $g^2$, which
provides accurate extrapolations to SU($N=\infty$)
\cite{mt3d}.
There also exist calculations of $T_c$ for
$D=2+1$ SU(2), SU(3) and SU(4) gauge theories 
\cite{Tc3d}.

For the sake of clarity I have relegated 
to appropriate Appendices the details of what I have done, and 
what lattice calculations I have used, in order to obtain the final 
continuum numbers that I quote in Section 2. I have tried to perform
the continuum extrapolations in a coherent way throughout
and I apologise to those authors who may feel that I have not 
done their calculations full justice in the process. For example, 
calculations with improved actions on coarse lattices 
raise particular difficulties; and, whenever in doubt, I have chosen 
to err on the conservative side.

\section{SU(2) and SU(3) in  the continuum}
\label		{su2su3_cont}

The spectrum of SU($N$) gauge theories is composed of states
that are composed entirely of gluons and so it is usual to 
call the single particle states (including resonances) 
by the generic name `glueballs'. Their masses we will
denote by $m_G$ or by  $m_{J^{PC}}$ where  ${J^{PC}}$ denotes
the quantum numbers of the state. These theories appear to be 
confining at low temperature, with a linear potential at large
distances between static sources in the fundamental representation.
We label the corresponding string tension by $\sigma$.
While this characterises the large-$r$ part of the potential,
one can characterise its behaviour at intermediate
distances by the distance $r_0$ at which the force, $F$, has a 
particular value. It has become customary to use the
particular definition $r_0^2F(r_0)=1.65$ (which corresponds
to a value that can be calculated with precision on 
the lattice and which can be estimated with some reliability
from the observed spectrum of heavy quark systems).
At sufficiently high temperatures the theory must
become deconfined, and we label the temperature at
which this happens by $T_c$. At short distances the theory
becomes asymptotically free and the scale controlling
the running of the coupling can be chosen to be 
$\Lambda_{\overline{MS}}$. SU($N$) gauge fields in 3+1
dimensions possess a topological charge, $Q$, and
a correponding topological susceptibility
$\chi_t = \langle Q^2 \rangle / V$ in a space-time
volume $V$. There are other condensates, in particular
the gluon condensate, but we choose to stop here.

In a lattice calculation the overall scale is set by
the lattice spacing $a$ which is determined implicitly
by the choice made for the inverse bare coupling 
$\beta=2N/g^2(a)$. So, for example, a mass calculation
will produce a number for the quantity $am_G(a)$, where
the first $a$ tells us that the mass is in lattice units,
and the other $a$ indicates that the mass will be
afflicted by lattice discretisation errors. We can
rid ourselves of the former by calculating the
ratio of two such masses: 
$am_G(a)/am_G^{\prime}(a) = m_G(a)/m_G^{\prime}(a)$.
For small $a$ one can expand the lattice action in
powers of $a$ and hence show that the leading corrections
are
\cite{symanzik}:
\begin{equation}
{{m_G(a)} \over {m_G^{\prime}(a)}}
=
{{m_G(0)} \over {m_G^{\prime}(0)}}
+
c_1 (a\mu)^2 + O(a^4)
\label{B1}
\end{equation}
where any convenient mass scale $\mu$, such as $m_G$
itself, may be used and where the coefficients 
$c_1 \ldots$ are power series in the (bare) coupling.

In the results I present in this paper I have usually 
extrapolated lattice masses to the continuum limit
using the string tension 
\begin{equation}
{{m_G(a)} \over {\surd\sigma(a)}}
=
{{m_G(0)} \over {\surd\sigma(0)}}
+
c a^2\sigma + \ldots
\label{B2}
\end{equation}
since that quantity is known with some accuracy over a 
very wide range of lattice spacings. I could have 
used $r_0$ equally well and the reader can easily translate
from one continuum ratio to another using the values
I tabulate. ($r_0/a$ is more accurately calculable than
$a\surd\sigma$ and so is likely to become the scale of
choice in the future.)

I have attempted to include all the useful `modern' lattice
calculations of these physical quantities and have
extrapolated them to the continuum limit using  
either a simple $O(a^2)$ correction, as in eqn(\ref{B2}),
or using both $a^2$ and $a^4$ corrections. Details are
in the Appendices. 

\begin{table}
\begin{center}
\begin{tabular}{|l|c|c|}\hline
\multicolumn{3}{|c|}{$m_G/\surd\sigma$} \\ \hline
state & SU(2) & SU(3) \\ \hline
$0^{++}$         & 3.74$\pm$0.12 & $3.64\pm0.09$ \\
$0^{++\ast}$     &               & $6.58\pm0.36$ \\
$2^{++}$         & 5.62$\pm$0.26 & $5.17\pm0.17$ \\
$2^{++\ast}$     &               & $6.73\pm0.73$ \\
$0^{-+}$         & 6.53$\pm$0.56 & $4.97\pm0.57$ \\
$2^{-+}$         & 7.46$\pm$0.50 & $7.05\pm0.65$ \\
$1^{++}$         & 10.2$\pm$0.5  &  \\
$1^{+-}$         &               & $6.32\pm0.31$ \\
$1^{-+}$         & [10.4$\pm$0.7] &  \\
$3^{++}$         & 9.0$\pm$0.7    &  \\
$3^{-+}$         & [9.8$\pm1.4$]  &  \\ \hline
\end{tabular}
\caption{\label{table_mk_su2su3}
Glueball masses, in units of the string tension, in the continuum
limit. Values in brackets have been obtained by extrapolating
from only two lattice values and so should be treated with caution.}
\end{center}
\end{table}
\nopagebreak
\begin{table}
\begin{center}
\begin{tabular}{|l|c|c|}\hline
\multicolumn{3}{|c|}{$m_G/m_{0^{++}}$} \\ \hline
state & SU(2) & SU(3) \\ \hline
$0^{++\ast}$     &               & $1.78\pm0.12$ \\
$2^{++}$         & $1.46\pm0.09$ & $1.42\pm0.06$ \\
$2^{++\ast}$     &               & $1.85\pm0.20$ \\
$0^{-+}$         & $1.78\pm0.24$ & $1.34\pm0.18$ \\
$2^{-+}$         & $2.03\pm0.20$ & $1.94\pm0.20$ \\
$1^{++}$         & $2.75\pm0.15$ &  \\
$1^{+-}$         &               & $1.73\pm0.09$ \\
$1^{-+}$         & [$3.03\pm0.31$] &  \\
$3^{++}$         & $2.46\pm0.23$ &  \\
$3^{-+}$         & [$2.91\pm0.47$] &  \\ \hline
\end{tabular}
\caption{\label{table_mm_su2su3}
Glueball masses, in units of the lightest scalar glueball mass,  
in the continuum limit.
Values in brackets have been obtained by extrapolating
from only two lattice values and so should be treated with caution.}
\end{center}
\end{table}

\begin{table}
\begin{center}
\begin{tabular}{|r|c|c|}\hline
     & SU(2) & SU(3) \\ \hline \hline
$r_0 \surd\sigma$   & $1.201\pm0.055$ & $1.195\pm0.010$ \\ \hline
$T_c/\surd\sigma$   & $0.694\pm0.018$ & $0.640\pm0.015$ \\ \hline
$\Lambda_{\overline{MS}}/\surd\sigma$   &  & $0.531\pm0.045$ \\ \hline
$\chi^{1\over4}_t/\surd\sigma$   & $0.486\pm0.010$ & $0.454\pm0.012$ \\ \hline
\end{tabular}
\caption{\label{table_rest_su2su3} The scale $r_0$, the deconfining
temperature, $T_c$, the topological susceptibility,
$\chi^{1\over4}_t$, and the (zero-flavour) $\Lambda$ parameter in
the ${\overline{MS}}$ scheme. All in the continuum and
in units of the string tension.} 
\end{center}
\end{table}

In Table~\ref{table_mk_su2su3} I list the continuum
glueball masses (in units of the string tension)
that I obtain, in this way, for both SU(2) and SU(3).
Another way to present this `data' is as a ratio of glueball masses.
This may be convenient for people who don't calculate the string
tension. I present the masses in this way in  
Table~\ref{table_mm_su2su3}. (For reasons explained in the 
Appendix these masses are not precisely what one would obtain
by dividing the appropriate entries in  Table~\ref{table_mk_su2su3}.)

So much for what is known about the mass spectrum. 
In  Table~\ref{table_rest_su2su3} I present the
continuum values of the other promised quantities.

The error estimates in these Tables are supposed to be realistic.
They include, albeit often indirectly, not only the statistical 
errors but also some of
the important systematic errors. For example, in the mass 
calculations sufficiently large lattices are used for the
finite-volume corrections to be very small. 
Using large lattices is slower and so leads to larger
statistical errors than using small lattices. Similarly 
the process of extrapolating to the continuum limit with one or
two correction terms magnifies the statistical errors on the
component lattice calculations. In this sense it would be
misleading to say that the errors in the Table are 
`only statistical'. The precautions taken during the 
calculation to control systematic errors typically
enlarge the statistical errors so that one should think of
the former as being included, albeit imperfectly, within the quoted 
errors. Perhaps the main caveat here concerns the identification
of the value of the spin $J$. A given irreducible representation
of the lattice rotational group will contain states with 
different values of $J$ in the continuum limit. One usually
labels the lattice masses by the minimum such value of $J$.
For calculations using operators smeared over physical
length scales this can usually be justified 
\cite{mt3d}
for the lightest states with given quantum numbers 
(most clearly for $P=+$). 
Thus nearly all the spin assignments in the Tables
should be correct. Nonetheless this still 
leaves open the possibility that, for example,
what we have called the $0^{++\star}$ is in reality
the $4^{++}$. In principle it is straightforward
to deal with this ambiguity
\cite{allJ}
but, for now, the ambiguity remains. Apart from this caveat,
it should be the case that the above 
numbers will withstand the test of time; at least within,
say, 2 of the quoted standard deviations.

\section{SU(2) and SU(3) at finite lattice spacing}
\label		{su2su3_latt}

The most comprehensive SU(3) lattice spectrum calculation 
is to be found in 
\cite{ukqcd64}.
(This title will be usurped by
\cite{MPnew}
once the latter ceases to be preliminary.)
The calculations were on a $32^3 64$ lattice at $\beta=6.4$,
which corresponds to a lattice spacing 
$a \simeq 0.1216/\surd\sigma \sim 0.05 fm$.
Similarly, in SU(2) there is a calculation
\cite{ukqcd285}
on a $48^3 56$ lattice at $\beta=2.85$ which 
corresponds to a lattice spacing 
$a \simeq 0.0636/\surd\sigma \sim 0.03 fm$.
The values in physical units come from
using $\surd\sigma \sim 450MeV$
\cite{MTnewton}, 
and are introduced
solely to make the point that these are very small
lattice spacings; they should not be taken more seriously 
than that.

We present the mass ratios from these papers in
Table~\ref{table_mk_su2su3_latt}. Since the lattice 
spacings are very small it is reasonable to
expect that the $O(a^2)$ lattice spacing corrections 
to the continuum values will also be small. However
the reader should be aware that this represents an
extra unquantified systematic error. It is clear that
in those cases where the states also appear in 
Table~\ref{table_mk_su2su3}, the reader should use 
the latter values.

\begin{table}
\begin{center}
\begin{tabular}{|c|c|c|}\hline
\multicolumn{3}{|c|}{$m_G/\surd\sigma$} \\ \hline
state & SU(2) ; $a \simeq 0.03 fm$ & SU(3) ; $a \simeq 0.05 fm$ \\ \hline
$0^{++}$   & $3.60\pm0.20$ & $3.52\pm0.12$ \\
$2^{++}$   & $5.55\pm0.27$ & $5.16\pm0.21$ \\
$1^{++}$   & $9.29\pm0.92$ & $9.0\pm0.7$ \\
$3^{++}$   & $9.75\pm0.85$ & $8.9\pm1.1$ \\ \hline
$0^{-+}$   & $7.04\pm0.57$ & $5.3\pm0.6$ \\
$2^{-+}$   & $7.70\pm0.50$ & $6.9\pm0.4$ \\
$1^{-+}$   & $10.3\pm0.9$ &  \\
$3^{-+}$   & $9.5\pm1.7$ &  \\ \hline
$2^{+-}$   &  & $7.7\pm1.0$ \\
$1^{+-}$   &  & $6.6\pm0.6$ \\ \hline
$2^{--}$   &  & $8.9\pm0.8$ \\
$1^{--}$   &  & $9.9\pm1.1$ \\ \hline
\end{tabular}
\caption{\label{table_mk_su2su3_latt}
Glueball masses in units of the string tension for fixed
but small values of the lattice spacing, as described in the 
text. No SU(3) estimates are given for the $0^{\pm-}, 3^{\pm-}, 
1^{-+}, 3^{-+}$
glueballs which appear to be too heavy to give a useful signal.}
\end{center}
\end{table}

The reason that our knowledge of the glueball mass spectrum 
is relatively incomplete is that since the
late 1980's there have been almost no dedicated glueball 
mass calculations; the few that have appeared have 
usually come as a side-product of other calculations.
Fortunately this is now changing and we can hope to
have accurate values for all the above states (and more)
quite soon. For a fore-taste see
\cite{MPnew},
although these are still `preliminary' and so have not been
included in this review.

\section{SU(4) and large N}
\label		{su4}

We see from Tables~\ref{table_mk_su2su3}-~\ref{table_mk_su2su3_latt}
that the physical  properties of SU(2) and SU(3) 
gauge theories are 
very similar (apart from the absence of any $C=-$ states
in the former case). It is tempting to see this as 
evidence that we are already close to the $N = \infty$ 
limit even with $N=2,3$. Since the leading corrections are
expected to be $O(1/N^2)$ this suggestion is not
absurd; and indeed it is exactly what is found to
occur for SU($N$) theories in 2+1 dimensions
\cite{mt3d}. 
If we assume that this is the case then we can 
extrapolate mass ratios to the $N=\infty$ limit using
\begin{equation}
{m \over {\surd\sigma}}\Biggl/_{\!N}
=
{m \over {\surd\sigma}}\Biggl/_{\!\infty}
+
 \ {c\over{N^2}}
\label{C1}
\end{equation}
and then we obtain the values displayed in 
Table~\ref{table_all_suN}.

\begin{table}
\begin{center}
\begin{tabular}{|r|c|}\hline
\multicolumn{2}{|c|}{$SU(\infty)$} \\ \hline
$m_{0^{++}}/\surd\sigma$     &   $3.56\pm0.18$   \\
$m_{2^{++}}/\surd\sigma$     &   $4.81\pm0.35$   \\
$T_c/\surd\sigma$   &   $0.597\pm0.030$  \\
$\chi^{1\over4}_t/\surd\sigma$   & $0.428\pm0.022$ \\
$r_0 \surd\sigma$   &   $1.190\pm0.043$  \\ \hline
\end{tabular}
\caption{\label{table_all_suN}
Various continuum mass ratios after extrapolation to $N=\infty$
from $N=2$ and $N=3$;
$assuming$ the validity of eqn(\ref{C1}) all the
way down to $N=2$. Errors correspond to $\chi^2=1$.}
\end{center}
\end{table}

Of course, since the fits have no degrees of freedom, we have 
no information on the reliability of using  eqn(\ref{C1}) all 
the way down to SU(2), and the numbers in Table~\ref{table_all_suN} 
must remain speculative. Clearly we need calculations for 
at least one further value of $N$; in that case we could 
test the goodness of the fits based on eqn(\ref{C1}). 
Unfortunately, while some exploratory SU(4) calculations 
do exist
\cite{mtsu4},
as summarised in Table~\ref{table_data_su4}, they 
are not accurate enough to permit a useful continuum
extrapolation. However all is not lost: 
the large-$N$ arguments apply not only
to the continuum limit but also to the theory at finite
values of the lattice spacing. This enables us to squeeze
some interesting information from Table~\ref{table_data_su4} 
as we shall now see.

\begin{table}
\begin{center}
\begin{tabular}{|l|c|c|c|}\hline
\multicolumn{4}{|c|}{$SU(4)$} \\ \hline
     & $\beta=10.7$ & $\beta=10.9$ & $\beta=11.1$  \\ \hline
$a\surd\sigma$     & $0.296\pm0.015$  & $0.228\pm0.007$  & 
$0.197\pm0.008$   \\
$am_{0^{++}}$      & $0.98\pm0.17$  & $0.75\pm0.07$  
& $0.77\pm0.06$   \\  
$am_{0^{++\star}}$ & $1.54\pm0.44$  & $1.39\pm0.16$  
& $1.03\pm0.14$   \\ 
$am_{2^{++}}$      & $1.28\pm0.27$  & $1.21\pm0.11$  
& $1.04\pm0.12$   \\ \hline  
\end{tabular}
\caption{\label{table_data_su4}
SU(4) masses and string tensions calculated in
\cite{mtsu4}, 
on $10^4$, $12^4$ and $16^4$ lattices at $\beta=$10.7, 10.9
and 11.1 respectively.}
\end{center}
\end{table}

The question we want to ask is whether there is a smooth large-$N$
limit and whether this is obtained by varying the coupling as 
$g^2 \propto 1/N$. This is what one expects from the usual analysis 
\cite{suN}
of Feynman diagrams (and what one finds to occur 
in 2+1 dimensions
\cite{mt3d}).

Since the coupling runs the latter part of the question
needs some elaboration. The usual $\beta$-function 
will ensure that $g^2 \propto 1/N$ for large $N$ provided 
that $\Lambda_{\overline{MS}}$ becomes independent of $N$ 
when expressed in units of a physical
length scale in the SU($N$) theory, such as $\surd\sigma$.
Equivalently suppose we define the running coupling, $g_N^2(l)$,
on a length scale $l \equiv c_l \xi_\sigma$ where
$\xi_\sigma \equiv 1/\surd\sigma$ and $c_l$ is small 
enough for the coupling to be small. Then what we are
looking for is
\begin{equation}
g_N^2(c_l \xi_\sigma) = 
{{N_0}\over{N}} g_{N_0}^2(c_l \xi_\sigma) +
O\biggl({1\over N^2}\biggr)
\label{C2}
\end{equation}
for all $N$ larger than some $N_0$. At the same time we
want physical mass ratios, bar some obvious exceptions, 
to have finite non-zero limits as $N\to\infty$.

We can test eqn(\ref{C2}) using the SU(4) string tensions
in Table~\ref{table_data_su4}. We note that the bare
coupling can be used to define a running coupling on
the scale of the cut-off $a$: $g^2(a) = 2N/\beta$.
However it is well-known that this is a `bad' way
to define a coupling and that one can do much better
with a mean-field 
\cite{parisi}
or tadpole-improved
\cite{lepage}
coupling defined by
\begin{equation}
g_I^2(a) = 
{{2N}\over{\beta \times {1\over N}\langle {\rm Tr} U_p \rangle}}
\label{C3}
\end{equation}
where $\langle {\rm Tr} U_p \rangle$ is the average plaquette.
We can extract three SU(4) couplings in this way 
corresponding to the three values of $\beta$ in
Table~\ref{table_data_su4}. The scales on which these couplings
are defined are given by the corresponding values of $a\surd\sigma$.
We now go to the SU(2) and SU(3) lattice string tension
values tabulated in the Appendices 
and find (by interpolation) the
mean-field improved lattice couplings corresponding to  
these three values of $a\surd\sigma$. We wish to test
the hypothesis that $Ng^2$ is becoming independent of $N$ 
at large $N$, so we list these couplings, multiplied by $N$,
in Table~\ref{table_g_suN}. We observe that these values
are indeed consistent with being independent of $N$. 
This provides some evidence that eqn(\ref{C2}) is indeed 
valid all the way down to SU(2). Presumably the $O(1/N)$ 
corrections would become visible in more accurate calculations.

\begin{table}
\begin{center}
\begin{tabular}{|c||c|c|c|}\hline
\multicolumn{4}{|c|}{$N g_I^2(l)$} \\ \hline
 $c_l=a\surd\sigma$    &  SU(2) & SU(3) & SU(4) \\ \hline
$0.197\pm0.008$  & $4.966\pm0.050$  & $4.938\pm0.045$  & 
4.943   \\
$0.228\pm0.007$  & $5.116\pm0.036$  & $5.097\pm0.036$  &
5.149   \\
$0.296\pm0.015$  & $5.434\pm0.076$  & $5.397\pm0.063$  &  
5.396   \\ \hline
\end{tabular}
\caption{\label{table_g_suN}
Bare couplings multiplied by $N$ on three different length scales 
$l=c_l\xi_\sigma$, each scale being the same for the SU(2), SU(3) 
and SU(4) gauge theories when expressed in physical units 
(here the string tension).}
\end{center}
\end{table}

Of course this analysis only makes sense if the theory
has a smooth limit at large $N$. We can investigate this
at each value of $a$ separately. In Table~\ref{table_0pp_suN}
we plot the values of $m_{0^{++}}/\surd\sigma$ for
$N=2,3,4$ at each value of $a\surd\sigma$. In
Table~\ref{table_2pp_suN} we do the same for 
$m_{2^{++}}/\surd\sigma$. We observe that these ratios
are indeed consistent with having a smooth, confining large-$N$
limit although the errors are too large for this
to be a taxing test. 

We conclude that there is significant evidence that
D=3+1 SU($N$) gauge theories are `close' to SU($\infty$)
all the way down to SU(2). However it is also clear
that very much better SU(4) calculations are needed 
(and indeed are in progress).  

\begin{table}
\begin{center}
\begin{tabular}{|c||c|c|c|}\hline
\multicolumn{4}{|c|}{$m_{0^{++}}/\surd\sigma$} \\ \hline
 $a\surd\sigma$    &  SU(2) & SU(3) & SU(4) \\ \hline
$0.197\pm0.008$ & $3.67\pm0.10$ & $3.33\pm0.05$ & $3.91\pm0.33$ \\
$0.228\pm0.007$ & $3.62\pm0.07$ & $3.22\pm0.04$ & $3.28\pm0.30$ \\
$0.296\pm0.015$ & $3.49\pm0.03$ & $2.94\pm0.08$ & $3.31\pm0.57$ \\
\hline
\end{tabular}
\caption{\label{table_0pp_suN}
The scalar glueball as a function of $N$, for
lattice spacings that are independent of $N$,
in units of the string tension.}
\end{center}
\end{table}

\begin{table}
\begin{center}
\begin{tabular}{|c||c|c|c|}\hline
\multicolumn{4}{|c|}{$m_{2^{++}}/\surd\sigma$} \\ \hline
 $a\surd\sigma$    &  SU(2) & SU(3) & SU(4) \\ \hline
$0.197\pm0.008$ & $5.59\pm0.09$ & $5.05\pm0.03$ & $5.23\pm0.61$ \\
$0.228\pm0.007$ & $5.58\pm0.03$ & $5.02\pm0.04$ & $5.31\pm0.48$ \\
$0.296\pm0.015$ & $5.54\pm0.14$ & $4.92\pm0.21$ & $4.30\pm0.90$ \\
\hline
\end{tabular}
\caption{\label{table_2pp_suN}
The tensor glueball as a function of $N$, for
lattice spacings that are independent of $N$,
in units of the string tension.}
\end{center}
\end{table}

\section{Comments and Prospects}
\label		{com}

Although it might seem, from Table~\ref{table_mk_su2su3}, that 
we already have a great deal of information about the
mass spectra of SU(2) and SU(3) gauge theories, this 
appearance is deceptive. As soon as one tries to match
these spectra with predictions of theoretical calculations
(for example 
\cite{dalley,AdS})
or models (for example
\cite{ip}),
one soon realises that it is important to have predictions 
for several excitations of the same quantum numbers as well 
as for spins which are different in the continuum but which
appear in the same representation of the lattice symmetry
group. For example, $J=0$ and $J=4$. Thus one needs
to develop techniques 
\cite{allJ} 
that will resolve such different continuum spins,
and one needs to be able to calculate accurately
heavier excited states. The latter will clearly entail developing
techniques for dealing with resonances and continuum states.

In order to take the continuum limit one needs a substantial range 
of lattice spacings, $a$, for which accurate calculations are 
available; extending from relatively coarse $a$ to as fine
$a$ as one can go. At the coarser end, the masses of 
heavier excited states are
so large in lattice units, $am_G$, as to be virtually
incalculable. Here the use of actions with different
timelike and spacelike lattice spacings will be crucial:
$a_s$ can be large while $a_tm_G$ remains calculably small
\cite{myasat,MPnew}.  
As we have seen in Appendix B, given the statistical
accuracy of such calculations, it is important that
the renormalisation of the asymmetry parameter be determined 
explicitly. Since part of the calculation must be performed 
for coarse $a$, there is an argument for using improved
actions which have smaller lattice discretisation errors.
If one does so then one might be tempted to perform
the whole calculation with coarse lattice spacings.
However we have seen that even if such actions reduce
the leading $O(a^2)$ errors, it is dangerous to assume
that these are so small as to be negligible compared to
the sub-leading $O(a^4)$ errors. That is to say, both
terms need to be included in the continuum extrapolation. 
In that case it is useful to have calculations at
smaller values of $a$ to accurately fix the coefficients
of such fits. 

In addition to mass spectra, we have reviewed calculations
of several other interesting physical quantities. Here
too there are questions that need answering. Most
pressing is a convincing demonstration that the
effective string theory that governs the long-distance
physics of the confining `flux tube' is indeed in the
universality class of the simple bosonic string (leading to
the usual L\"uscher correction in eqn(\ref{AA1}) and the
corresponding correction for the static potential).
Whether this is in fact so depends on the effective degrees
of freedom (central charge) of the effective string
theory. There is numerical evidence for the usual
assumption but it is not yet accurate enough to be
really convincing. Also needed is a resolution of
the $\sim 4\%$ discrepancy between the two groups of 
calculations of the SU(3) deconfining temperature, as
discussed in Appendix B.

A major gap at present is the lack of an accurate
SU(4) calculation. While we have managed to squeeze
a surprising amount out of the crude lattice calculations
\cite{mtsu4}
that are currently available, a proper quantitative
discussion of the large-$N$ limit in 3+1 dimensions
must await a much better set of calculations.

\vspace{0.5cm}

%
%
\noindent { {\bf Acknowledgements:}}
I have benefitted from discussions with many of the
authors I quote. I am grateful to Rainer Sommer
for providing me with the unpublished SU(2) $r_0$ values. 

\newpage

\noindent{\Large{\bf Appendix A : SU(2) }}

\vspace{0.45cm}

In this Appendix I describe in some detail how the SU(2)
continuum numbers in the main body of the text were
arrived at. 

\vspace{0.35cm}

\noindent {\large {\bf  (a) the mass spectrum and string tension}}

\vspace{0.35cm}

The most extensive calculations are for the string tension and the
lightest scalar and tensor glueballs. The values that I have used
for my continuum extrapolations are listed in
Table~\ref{table_02K_su2}. There are also some calculations for
other glueball quantum numbers and these are listed
in Table~\ref{table_mrest_su2}. All these calculations use
the standard Wilson plaquette action
\cite{wilson}.
The sources for these calculations are as follows: 
$\beta=2.0$ and $\beta=2.1$ are from
\cite{mtunpub};
$\beta=2.2$ to $\beta=2.6$ are from
\cite{cmmtsu2};
except for the $32^4$ lattice at $\beta=2.5$ which comes from
\cite{cmsu2}
as does the $\beta=2.7$ calculation; $\beta=2.85$ is from
\cite{ukqcdsu2}.
There are some other string tension calculations that I have
not used. Notably 
\cite{balisu2}
where one finds $a\surd\sigma =0.1871(10),0.1871(32)$
on $16^4, 32^4$ lattices at $\beta=2.5$; 
$a\surd\sigma =0.1207(3)$ on a $48^3 64$ lattice at
$\beta=2.635$; and $a\surd\sigma = 0.0911(4)$ on a 
$32^4$ lattice at $\beta=2.74$. (This last lattice is
probably too small, in physical units, to be
entirely trustworthy.) These values are consistent 
with the ones we list.

Some technical asides. The values listed are effective masses from 
$t=a$ to $t=2a$ for $\beta\leq 2.4$ for the string tension,
for $\beta\leq 2.5$ for the $J=0,2$ glueballs, and for all 
$\beta$ for the $J=1,3$ glueballs. For other (higher) $\beta$
I list effective masses from $t=2a$ to $t=3a$. When I
need a single value of the string tension at $\beta=2.4$ 
or at $\beta=2.5$, I use $a\surd\sigma=0.2673(15)$ and 
$a\surd\sigma=0.186(3)$
respectively. The string tensions are all obtained from the 
masses of torelons. These are loops of fundamental flux closed 
through the periodic spatial boundary and hence neither 
contractible nor, in a confining phase, breakable. For
a loop of length $L$ in lattice units, one obtains $\sigma$ 
from the mass $m_l$ of the loop using
\begin{equation}
a m_l = a^2\sigma L  - {\pi \over {3 L}}.
\label{AA1}
\end{equation}
The (leading) correction in eqn(\ref{AA1}) 
is a universal string `Casimir energy', and
the coefficient used in eqn(\ref{AA1}) assumes that the 
effective confining string belongs to
the simplest bosonic string universality class
\cite{stringcor}. 
There is some numerical evidence for this apparently
reasonable assumption but it is not so precise
as to be convincing. This correction typically contributes
$\sim 4\%$ to the values in  Table~\ref{table_02K_su2}.

\begin{table}[tbh]
\begin{center}
\begin{tabular}{|c|c|c|c|c|c|c|}\hline
\multicolumn{7}{|c|}{SU(2)} \\ \hline
 $\beta$  & lattice & $a\surd\sigma$ & $am_{0^{+}}$ & $am_{2^{+}}$  
& $am_{0^{-}}$ & $am_{2^{-}}$\\ \hline
 2.0  & $6^3 12$ & 0.69(5)    & 1.76(16)  & & & \\
 2.1  & $6^3 12$ & 0.608(16)  & 1.74(13)  & & & \\
 2.2  & $8^3 10$ & 0.467(10)  & 1.44(8)   & 2.76(45) & & \\
 2.3  & $10^4$   & 0.3687(22) & 1.216(23) & 1.94(8)  & & \\
 2.4  & $14^4$   & 0.2686(21) & 0.959(20) & 1.487(33)& & \\
 2.4  & $16^4$   & 0.2660(21) & 0.936(23) & 1.520(30)  & & \\
 2.5  & $16^4$   & 0.1834(26) & 0.660(22) & 1.028(27)& &\\
 2.5  & $20^4$   & 0.1881(28) & 0.723(23) & 1.051(26)& 
1.241(63) & 1.34(13) \\
 2.5  & $32^4$   & 0.187(12)  & 0.70(4)   & 0.99(4)  & 
1.37(14)  & 1.42(11) \\
 2.6  & $20^4$   & 0.1326(30) & 0.51(3)   & 0.73(4)  & & \\
 2.7  & $32^4$   & 0.1005(20) & 0.39(3)   & 0.55(3)  & 
0.63(6)   & 0.72(4)  \\
 2.85 & $48^3 56$ & 0.0636(12) & 0.229(12) & 0.353(16) & 
0.430(35) & 0.49(3)  \\ \hline
\end{tabular}
\caption{\label{table_02K_su2}
Various SU(2) lattice calculations of the string tension and
lightest $J=0$ and $J=2$ masses. Errors in brackets.}
\end{center}
\end{table}

\begin{table}[tbh]
\begin{center}
\begin{tabular}{|c|c|c|c|c|c|}\hline
\multicolumn{6}{|c|}{SU(2)} \\ \hline
 $\beta$  & lattice & $am_{3^{+}}$ & 
$am_{3^{-}}$ & $am_{1^{+}}$ & $am_{1^{-}}$ \\ \hline
 2.5  & $32^4$   & 1.88(43) & & 1.70(20) & \\
 2.7  & $32^4$   & 0.90(5) & 1.32(11) & 1.02(3) & 1.11(4) \\
 2.85 & $48^3 56$ & 0.588(36) & 0.71(5) & 0.626(27) 
& 0.677(21) \\ \hline
\end{tabular}
\caption{\label{table_mrest_su2}
Various SU(2) lattice calculations of the
lightest $J=1$ and $J=3$ masses. Errors in brackets.}
\end{center}
\end{table}

To obtain the continuum limits listed in 
Table~\ref{table_mk_su2su3}
I have taken the lattice values in the above Tables 
and used eqn(\ref{B2}).
If the fit is bad then the lowest-$\beta$ values are
dropped till a good fit is obtained. I use conservative 
$25\%$ confidence level (CL) limits for the errors;
less where the best fit has $CL \leq 70\%$. In the case of
the $0^+$ and of the $2^+$ one can fit all the way down to
quite coarse lattice spacings ($\beta=2.0$ and $\beta=2.2$
respectively) and so there I also fit with
an extra $O(a^4)$ correction to check that the resulting
continuum extrapolations are consistent with the values I
quote. I follow a similar procedure to obtain the mass
ratios in  Table~\ref{table_mm_su2su3}.

In addition to the above calculations, which use the Wilson
plaquette action, there are a number of calculations
using other actions. These provide a useful universality 
check and I list in  Table~\ref{table_mimp_su2} my
extrapolations to the continuum limit of these lattice
calculations. The first three calculations are from
\cite{trottier}. 
All three use an anisotropic lattice action with 
the ratio of timelike and spacelike lattice spacings
chosen to be small: $\xi\equiv a_t/a_s \ll 1$.
This is a trick 
\cite{myasat}
that enables one to work with coarse $a_s$ and yet have 
masses, $a_tm_G$, that are not so heavy as to be incalculable.
The first two actions are variants of tadpole-improvement
\cite{lepage}
while the third is an (anisotropic) Wilson action.
The calculations take advantage of the improvement by
being performed on quite coarse lattices. For this
reason we have extrapolated these results to the 
continuum with a more complicated $c_1a^2\sigma+c_2(a^2\sigma)^2$
correction term. A potential problem 
with these anisotropic lattice calculations is that they
have been performed with values of $\xi$ that vary strongly
with $\beta$. This makes it tricky to motivate a simple
uniform continuum extrapolation. The Manton
\cite{manton}
and Symanzik tree-level improved action
\cite{symanzik} 
calculations are from
\cite{cmmtsu2}.
These calculations suffer from having been performed 
at only two values of $\beta$, but these are large enough
that a simple $O(a^2)$ correction should suffice.
The errors on the extrapolations 
correspond to $\chi^2=1$ and are not really reliable. 
Despite these caveats, it is clear
that we have significant confirmation of universality from
all these results.

\begin{table}[tbh]
\begin{center}
\begin{tabular}{|c|c|c|c|c|}\hline
\multicolumn{5}{|c|}{SU(2) -- improved actions} \\ \hline
 action  & $m_{0^{+}}/\surd\sigma$ &  $m_{2^{+}}/\surd\sigma$ & 
$m_{0^{-}}/\surd\sigma$ &  $m_{2^{-}}/\surd\sigma$ \\ \hline
tadpole1($a_s\not=a_t$) &  $3.97\pm0.31$ & $5.0\pm1.1$   &  &  \\
tadpole2($a_s\not=a_t$) &  $4.4\pm0.5$   &  &  &  \\
Wilson($a_s\not=a_t$)   & $3.76\pm0.22$  & $5.91\pm0.44$ &  &  \\
Symanzik($a_s=a_t$)     & $3.93\pm0.21$  & $5.67\pm0.31$ & $6.77\pm0.53$ 
& $6.78\pm0.80$ \\
Manton($a_s=a_t$)       & $4.01\pm0.28$  & $5.72\pm0.25$ & $6.71\pm0.56$ 
& $7.49\pm0.60$ \\ \hline
\end{tabular}
\caption{\label{table_mimp_su2}
SU(2) mass spectra, in the continuum limit, from calculations using a 
variety of `improved' lattice actions (see text).}
\end{center}
\end{table}

\vspace{0.35cm}

\noindent {\large {\bf (b) the $\bf r_0$ parameter}}

\vspace{0.35cm}

The SU(2) $r_0$ parameter has been calculated for a range of
lattice spacings in
\cite{sommersu2}.
In Table~\ref{table_r0_su2} I present the raw numbers
\cite{sommerpc}
and the corresponding string tensions, which are taken from 
Table~\ref{table_02K_su2} either directly or by interpolation. 
One can extrapolate to the continuum limit using either
an $a^2\sigma$ or an $(a/r_0)^2$ correction. These give essentially
identical best fits: for example
\begin{equation}
{{r_0}{\surd\sigma}} = 1.201(55) - 0.56(2.48) a^2\sigma .
\label{AA2}
\end{equation}
with $80\%$ confidence level.

\begin{table}[tbh]
\begin{center}
\begin{tabular}{|c|c|c|}\hline
\multicolumn{3}{|c|}{SU(2)} \\ \hline
 $\beta$  & $r_0/a$ &  $a\surd\sigma$ \\ \hline
 2.50  & $6.39\pm0.09$ & $0.186\pm0.003$  \\
 2.55  & $7.55\pm0.11$ & $0.157\pm0.003$ \\
 2.60  & $8.82\pm0.08$ & $0.1326\pm0.0030$ \\
 2.70  & $12.05\pm0.22$ & $0.1005\pm0.0020$ \\
 2.85  & $19.08\pm0.88$ & $0.0636\pm0.0012$ \\ \hline
\end{tabular}
\caption{\label{table_r0_su2}
Values of the $r_0$ parameter and string tension.}
\end{center}
\end{table}

\vspace{0.35cm}

\noindent {\large {\bf (c) the deconfining temperature {$\bf T_c$}}}

\vspace{0.35cm}

A periodic Euclidean lattice of size $N^3_s N_\tau$
with $N_s \gg N_\tau$ may be used to calculate the
static properties of the gauge theory at a 
temperature  $T=1/aN_\tau$. In Table~\ref{table_Tc_su2} 
I list the critical coupling $\beta_c$ corresponding to the
SU(2) deconfining temperature for various $N_\tau$ (see
\cite{Tcsu2}
and references therein)
together with my interpolated values of the string tension.
(These values have been extrapolated to $N_s=\infty$ using
a finite-size scaling analysis.) Thus
\begin{equation}
T_c = {1\over{N_\tau a(\beta_c)}}.
\label{AA3}
\end{equation}
and we can extrapolate to the continuum limit with
\begin{equation}
{{T_c}\over{\surd\sigma}} = 0.694(18) - 0.28(33) (aT_c)^2
\label{AA4}
\end{equation}
or using a $a^2\sigma$ correction. The best fit has an
excellent $90\%$ confidence level.

\begin{table}[tbh]
\begin{center}
\begin{tabular}{|c|c|c|}\hline
\multicolumn{3}{|c|}{SU(2)} \\ \hline
 $N_\tau$  & $\beta_c$ &  $a\surd\sigma$ \\ \hline
 2   & $1.8800\pm0.0030$ &  \\
 3   & $2.1768\pm0.0030$ & $0.4950\pm0.0110$ \\
 4   & $2.2986\pm0.0006$ & $0.3704\pm0.0023$ \\
 5   & $2.3726\pm0.0045$ & $0.2920\pm0.0045$ \\
 6   & $2.4265\pm0.0030$ & $0.2435\pm0.0032$ \\
 8   & $2.5115\pm0.0040$ & $0.1787\pm0.0040$ \\
 16  & $2.7395\pm0.0100$ & $0.0891\pm0.0034$ \\ \hline
\end{tabular}
\caption{\label{table_Tc_su2}
Values of the deconfining critical coupling and string tension.}
\end{center}
\end{table}

\vspace{0.35cm}

\noindent {\large {\bf (d) the topological susceptibility {$\bf \chi_t$}}}

\vspace{0.35cm}

If $Q$ is the topological charge of a gauge field then the
susceptibility is defined as $\chi_t = \langle Q^2 \rangle/V$
where $V$ is the space-time volume. So on an $L^4$ lattice,
$a^4\chi_t = \langle Q^2 \rangle/L^4$. How to calculate the 
topological charge $Q$ of a lattice gauge field is a problem
that is now well-understood
\cite{MTnewton}.
Here we will summarise several calculations and use them to
extract a continuum value for $\chi^{1/4}_t/\surd\sigma$. Our
description of the methods used is inadequately brief 
and the reader
needs to consult the literature for answers to a variety of
natural questions.

The oldest reliable way to calculate the topological
charge $Q$ of a lattice 
gauge field is to locally smoothen the gauge field, for
example by minimising
its action density, and then to calculate a lattice topological
charge density $Q_L(x)\simeq a^4 Q(x)$ at every site, and finally
to sum that up over all lattice sites. This approach goes by the name
of `cooling'
\cite{MTcool,MTnewton}. I will call the cooled total charge 
$Q_c$. The fields will contain some charges that are
narrow and these will be particularly influenced by the 
presence of the lattice cut-off. So it is useful 
to define a charge $Q_{c^\prime}$ from which such narrow 
charges have been excluded. The corresponding two susceptibilites,
$\chi_c$ and $\chi_{c^\prime}$, will differ at finite-$a$
but should converge in the continuum limit. (Note that, for the sake 
of clarity, I drop the subscript $t$ on $\chi$.) We shall use a
cut-off on the instanton size that corresponds 
to all charges with a peak height
$Q_L(x_{peak}) \geq 1/16\pi^2$ being excluded. The
calculations I shall use are from
\cite{cmmtQ,mtdp,mtunpub}.
In these calculations the number of cooling sweeps
is typically 15 or 20. Since the topological charge
varies negligibly with cooling in this range, we can safely
average such slightly different calculations. A more 
recent `improved' cooling method, whose charges we
label by $Q_{Ic}$, is to be found in 
\cite{Zurichsu2}.
An approach using a renormalisation group inspired form
of cooling has been developed in
\cite{Bouldersu2}
and the corresponding charges we label $Q_{RG}$. 
(For the latter we shall
use the values after `0 smears'.) There also exist
methods that do not smoothen the gauge fields. In
particular one can, instead, form smooth smeared operators 
and apply them directly to the original gauge fields, 
correcting explicitly for the additive
and multiplicative renormalisations, as is done in 
\cite{Pisasu2}.
We shall refer to this susceptibility as $\chi_{sm}$.
(We use the values after 2 `smears'.) 
Finally one can use a geometric form of the total topological
charge
\cite{mtdp}. 
To avoid problems with dislocations this is
calculated on blocked lattice fields and averaged over
all 16 blockings. This charge we call $Q_G$. (As one takes $a\to 0$ the
number of blockings will, in principle, need to be
increased.) All these methods 
will give different results at a fixed value of $a$, because
they deal differently with instantons whose sizes are 
near the cut-off. However they should converge to the
same continuum value.

In Table~\ref{table_Q_su2} I list the 6 different
susceptibilities.
It is apparent that the variation between the
different calculations decreases with increasing $\beta$.
In Table~\ref{table_Qcont_su2} I give the continuum
extrapolation of $\chi^{1/4}_t/\surd\sigma$ for each of these
six calculations. (I have used an $a^2\sigma$ lattice
correction.) They are all pretty much consistent with each
other and the final value I infer, and quote in 
Table~\ref{table_rest_su2su3}, is an average value
$\chi^{1/4}_t/\surd\sigma=0.486(10)$.

\begin{table}[tbh]
\begin{center}
\begin{tabular}{|c|c|c|c|c|c|c|}\hline
\multicolumn{7}{|c|}{SU(2)} \\ \hline
$\beta$ & $a\chi^{1/4}_c$  & $a\chi^{1/4}_{c^\prime}$ 
& $a\chi^{1/4}_G$  & $a\chi^{1/4}_{Ic}$  
& $a\chi^{1/4}_{RG}$  & $a\chi^{1/4}_{sm}$  \\ \hline
 2.20   & 0.1481(53) &  &  &  &  &  \\
 2.30   & 0.1420(21) & 0.1024(25) & 0.1211(30) &  &  &  \\
 2.40   & 0.1163(13) & 0.0937(27) & 0.1026(25) & 0.1216(18) & 0.1299(21) &  \\
 2.44   &  &  &  &  &  & 0.1027(19) \\
 2.50   & 0.0844(11) & 0.0760(12) & 0.0800(18) & 0.0913(30) & 0.0957(21) &  \\
 2.5115 &  &  &  &  &  & 0.0820(11) \\
 2.57   &  &  &  &  &  & 0.0681(12  \\
 2.60   & 0.0627(13) & 0.0585(14) & 0.0610(16) & 0.0605(46) & 0.0674(15) &  \\ \hline
\end{tabular}
\caption{\label{table_Q_su2}
The topological susceptibility, using various methods as defined
in the text.}
\end{center}
\end{table}

\begin{table}[tbh]
\begin{center}
\begin{tabular}{|c|c|c|c|c|c|}\hline
\multicolumn{6}{|c|}{SU(2) : $\chi^{1/4}_n/\surd\sigma$} \\ \hline
$n=c$  & $n=c^\prime$ & $n=G$ & $n=Ic$  & $n=RG$  & $n=sm$  \\ \hline
 0.486(12) &  0.468(15) & 0.480(18) & 0.501(45) & 0.528(21) & 0.480(23) \\
\hline
\end{tabular}
\caption{\label{table_Qcont_su2}
The continuum topological susceptibility, from the values in
Table~\ref{table_Q_su2}, in units of the string  tension.}
\end{center}
\end{table}

\vspace{1.0cm}

\noindent{\Large{\bf Appendix B : SU(3) }}

\vspace{0.45cm}
In this Appendix I will describe the SU(3) lattice calculations
which I use to obtain the continuum numbers in the main body of
this paper.

\vspace{0.35cm}

\noindent {\large {\bf (a) the mass spectrum and the string tension}}

\vspace{0.35cm}

In Table~\ref{table_K_su3} I present the lattice calculations
of the string tension that I use. Where there is more than
one reference listed, I have taken an average of the
different values (checking, of course, that they 
are consistent with each other).  
For later use, values of the string tension have also
been estimated by interpolation for a variety of
intermediate $\beta$ values at which there are 
no direct calculations.

\begin{table}[tbh]
\begin{center}
\begin{tabular}{|c|c|c|}\hline
\multicolumn{3}{|c|}{SU(3)} \\ \hline
$\beta$  & $a\surd\sigma$ & sources \\ \hline
 5.50  &  $0.5830\pm0.0130$  & \cite{fsstK} \\
 5.54  &  $0.5727\pm0.0052$  & \cite{ehkr0} \\
 5.60  &  $0.5064\pm0.0028$  & \cite{ehkr0} \\
 5.70  &  $0.3879\pm0.0039$  & \cite{ehkr0} \\
 5.80  &  $0.3176\pm0.0016$  & int \\
 5.83  &  $0.2991\pm0.0009$  & int \\
 5.85  &  $0.2874\pm0.0007$  & \cite{ehkr0} \\
 5.90  &  $0.2613\pm0.0028$  & \cite{cmmtsu3} \\
 5.93  &  $0.2485\pm0.0012$  & int \\
 5.95  &  $0.2391\pm0.0014$  & int \\
 6.00  &  $0.2188\pm0.0008$  & \cite{ehkr0,cmmtsu3,spcmsu3,bs2,ukqcd3} \\
 6.07  &  $0.1964\pm0.0010$  & int \\
 6.10  &  $0.1876\pm0.0012$  & int \\
 6.17  &  $0.1684\pm0.0012$  & int \\
 6.20  &  $0.1608\pm0.0010$  & \cite{cmmtsu3,spcmsu3,bs1,ukqcd1,ukqcd3} \\
 6.30  &  $0.1398\pm0.0012$  & int \\
 6.40  &  $0.1216\pm0.0011$  & \cite{ukqcd64,bs2,ukqcd3} \\
 6.50  &  $0.1068\pm0.0009$  & \cite{ukqcd2} \\
 6.57  &  $0.0975\pm0.0012$  & int \\ \hline
\end{tabular}
\caption{\label{table_K_su3}
The string tension with sources (`int' means obtained
by interpolation).}
\end{center}
\end{table}

The calculations of the SU(3) mass spectrum fall into 
two groups which are sufficiently different that they
need separate treatment. The first and older category, 
involves calculations that use the standard plaquette action 
and are performed down to small values of the lattice spacing $a$.
These allow accurate continuum extrapolations
for the lightest glueballs. For the heavier glueballs
they are much poorer. This is partly because many of the earlier
calculations did not attempt to calculate them and, as a result,
there is often too short a lever arm for a controlled continuum 
extrapolation. In addition heavy glueballs have larger
values of $am_G$ and so the correlation function will
often drop into the statistical noise before an effective
mass plateau is really clear -- except at the very smallest
values of $a$, where the lattices have to be very large
and hence computationally expensive. 

The second category of calculations is recent
and uses improved lattice actions so that useful
calculations can be performed for coarse lattice spacings.
To avoid large values of  $am_G$ and the consequent
danger of having no useful signal,
they use an anisotropic action where the space and time 
lattice spacings differ and satisfy $a_t \ll a_s$. 
In that case one can have a small and easily measured
value of $a_tm_G$, even when the glueball is heavy, while 
at the same time $a_s$ is large so that the spatial lattice can 
be small. I will now discuss these two kinds of calculations 
with $a_s=a_t$ and $a_s\not=a_t$, in turn.

\vspace{0.15cm}

\noindent {$\bullet \ \ \ ${\bf $\bf a_s=a_t$ calculations}}

\vspace{0.15cm}

The lattice mass values that I use are listed in 
Tables~\ref{table_m02sym_su3},~\ref{table_msym_su3}.
This is an (almost) complete collection of `modern'
mass calculations where systematic errors should be
negligible. The calculations extend to small values of $a$,
as we can see by substituting $\surd\sigma \sim 0.5 fm$
in Table~\ref{table_K_su3}. Thus one expects that over
at least part of the range of $a$ in which calculations
exist, the leading $O(a^2)$
corrections should dominate and we can perform
extrapolations using
\begin{equation}
{{m_G(a)} \over {\surd\sigma(a)}}
=
{{m_G(0)} \over {\surd\sigma(0)}}
+
c a^2\sigma .
\label{BB1}
\end{equation}
We find that this functional form fits the 
$0^{++}$ lattice values for $\beta \geq 5.7$
and the $2^{++}$ lattice values for $\beta \geq 5.9$.
For the $0^{-+}$ and  $2^{-+}$ we include all the
lattice values in the fits, but for the  $1^{+-}$
we find that we have to drop the $\beta=5.9$ value.
For the  $0^{++\star}$ we have no degrees of freedom
so we have no feedback about the validity of the
fit in that case. The resulting continuum values are
shown, a little later, in Table~\ref{table_mcont_su3}.

\begin{table}[tbh]
\begin{center}
\begin{tabular}{|c|c|c|c|c|}\hline
\multicolumn{5}{|c|}{SU(3) : $am_G$} \\ \hline
$\beta$  & $0^{++}$ & $2^{++}$ & $0^{++\star}$ & sources \\ \hline
 5.50  &  $1.14\pm0.03$    &    &     &\cite{fsstM} \\
 5.70  &  $0.974\pm0.029$  &    &     &\cite{GF11} \\ 
 5.70  &  $0.90\pm0.04$    &    &     &\cite{fsstM} \\ 
 5.83  &  $0.858\pm0.043$  &    &     &\cite{GF11} \\  
 5.90  &  $0.809\pm0.024$  &  $1.342\pm0.039$  & & \cite{cmmtsu3} \\  
 5.93  &  $0.779\pm0.008$  &  $1.224\pm0.020$  & & \cite{GF11} \\  
 6.00  &  $0.710\pm0.022$  &  $1.11\pm0.03$    & $1.28\pm0.05$ & \cite{cmmtsu3} \\  
 6.17  &  $0.582\pm0.010$  &  $0.828\pm0.026$  & & \cite{GF11} \\  
 6.20  &  $0.560\pm0.020$  &  $0.845\pm0.020$  & $1.00\pm0.03$ & \cite{cmmtsu3} \\  
 6.40  &  $0.415\pm0.014$  &  $0.609\pm0.016$  & & \cite{ukqcd64} \\  
 6.40  &  $0.433\pm0.011$  &  $0.636\pm0.023$  & & \cite{GF11} \\ \hline  
\end{tabular}
\caption{\label{table_m02sym_su3}
The lightest $0^{++}$ and $2^{++}$ glueball masses and the 
first scalar excitation, using lattice actions with $a_s=a_t$.}
\end{center}
\end{table}

\begin{table}[tbh]
\begin{center}
\begin{tabular}{|c|c|c|c|c|}\hline
\multicolumn{5}{|c|}{SU(3) : $am_G$} \\ \hline
$\beta$  & $0^{-+}$ & $2^{-+}$ & $1^{+-}$ & sources \\ \hline
 5.90 & $1.65\pm0.20$ & $1.60\pm0.18$ & $1.90\pm0.20$  & \cite{cmmtsu3} \\  
 6.00 & $1.25\pm0.14$ & $1.28\pm0.15$ & $1.33\pm0.14$  & \cite{cmmtsu3} \\  
 6.20 & $0.92\pm0.08$ & $1.20\pm0.09$ & $0.984\pm0.045$ & \cite{cmmtsu3} \\  
 6.40 & $0.63\pm0.04$ & $0.81\pm0.04$ & $0.780\pm0.030$ &
\cite{ukqcd64} \\  \hline
\end{tabular}
\caption{\label{table_msym_su3}
The lightest $0^{-+}$, $2^{-+}$ and $1^{+-}$ glueball masses, 
using lattice actions with $a_s=a_t$.}
\end{center}
\end{table}

\vspace{0.15cm}

\noindent{$\bullet \ \ \ ${\bf $\bf a_s\not=a_t$ calculations}}

\vspace{0.15cm}

This type of calculation is very recent and so the
important questions have not all been resolved in
the way that they have for the older and now
well-established $a_s=a_t$ calculations. On the other
hand these calculations certainly embody features that
will be part of future mass calculations, so it is
worth discussing them in some detail.

At the moment there is only one complete published 
glueball spectrum calculation of this kind
\cite{MPasym}.
To be precise, this  paper contains two separate calculations, one
with a tree-level anisotropy of $\xi = (a_s/a_t)_{tree} = 3$ 
and a second with $\xi = (a_s/a_t)_{tree} = 5$. An improved
action is used which should reduce the $O(a_s^2)$ errors 
down to $O(\alpha_s a_s^2)$. The $O(a_t^2)$ errors should
be small because $a_t$ is (very) small, and so are ignored. 

The lattice masses we use here are precisely the ones
that are highlighted in the tables of 
\cite{MPasym}
and there is no point reproducing all those readily accessible
numbers here. However, to give the reader a flavour of their results
we provide in Table~\ref{table_masym3_su3} the lightest
$0^{++}$, $2^{++}$(=$E^{++}$) and $1^{+-}$ masses together with 
the scale $r_0$ in lattice units, all for the $\xi=3$ calculation. 
Also shown is the quantity
$r_0^2\sigma$ which, being a dimensionless ratio of physical
quantities, is in a different category from the other quantities 
listed. Table~\ref{table_masym5_su3} lists the same quantities for
the $\xi=5$ calculation. We note that by using $a_t\ll a_s$
it is possible to get accurate lattice mass estimates even
for the heavier states, despite the fact that the spatial
lattice spacing is large: $a_s \sim 0.2 - 0.4 fm$. 
If we use the listed values of $a_s/r_0$ and $r^2_0\sigma$ and
compare to the values in Table~\ref{table_K_su3} we see that
the smallest $a_s$ in these calculations corresponds to
$\beta \sim 5.60$ in the symmetric lattice calculations.
These are indeed $very$ coarse (spatial) lattices.

\begin{table}[tbh]
\begin{center}
\begin{tabular}{|c||c|c|c|c||c|}\hline
\multicolumn{6}{|c|}{SU(3) : $\xi=3$} \\ \hline
$\beta$  & $a_tm_{0^{++}}$ &  $a_tm_{2^{++}}$ &  $a_tm_{1^{+-}}$
& $a_s/r_0$ & $r^2_o\sigma$ \\ \hline
 1.7 & 1.05(2)  & 1.65(5)  & 2.09(9) & 0.861(2)  & 1.58(1)  \\
 1.9 & 0.87(1)  & 1.47(4)  & 1.80(6) & 0.773(2)  & 1.52(2)  \\
 2.0 & 0.797(9) & 1.40(2)  & 1.68(3) & 0.7271(8) & 1.462(7) \\
 2.2 & 0.649(8) & 1.19(2)  & 1.48(3) & 0.6192(8) & 1.362(8) \\
 2.4 & 0.548(6) & 0.995(9) & 1.24(1) & 0.505(1)  & 1.329(6) \\
 2.6 & 0.464(7) & 0.781(8) & 0.97(1) & 0.4021(9) & 1.340(2) \\ \hline
\end{tabular}
\caption{\label{table_masym3_su3}
The lightest $0^{++}$, $2^{++}(=E^{++})$ and $1^{+-}$ glueball masses, 
the parameter $r_0$ and its relation to the string tension,
all for the $\xi=(a_t/a_s)_{tree}=3$ calculation.}
\end{center}
\end{table}

\begin{table}[tbh]
\begin{center}
\begin{tabular}{|c||c|c|c|c||c|}\hline
\multicolumn{6}{|c|}{SU(3) : $\xi=5$} \\ \hline
$\beta$  & $a_tm_{0^{++}}$ &  $a_tm_{2^{++}}$  &  $a_tm_{1^{+-}}$
& $a_s/r_0$ & $r^2_o\sigma$ \\ \hline
 1.7 & 0.578(5) & 1.103(8) & 1.19(1)  & 0.8169(9) & 1.473(9) \\
 1.9 & 0.475(4) & 0.918(7) & 1.053(8) & 0.727(1)  & 1.45(1)  \\
 2.2 & 0.362(3) & 0.686(4) & 0.819(4) & 0.5680(5) & 1.356(4) \\
 2.4 & 0.303(3) & 0.542(2) & 0.652(5) & 0.459(1)  & 1.342(4) \\ \hline
\end{tabular}
\caption{\label{table_masym5_su3}
The lightest $0^{++}$, $2^{++}(=T^{++}_2)$ and $1^{+-}$ glueball masses, 
the parameter $r_0$ and its relation to the string tension,
all for the $\xi=(a_t/a_s)_{tree}=5$ calculation.}
\end{center}
\end{table}

There are at least two questions that need answering if we are to
be able to extract continuum numbers from these calculations:

\noindent{1) what is $\xi_r \equiv a_s/a_t$?}

\noindent{2) what are the lattice corrections to the continuum limit
of mass ratios?}

Let us start with the first question.
The tree level value of $a_s/a_t=\xi$ will be renormalised,
\begin{equation}
\biggl({{a_s}\over{a_t}}\biggr)
=
\biggl({{a_s}\over{a_t}}\biggr)_{tree}
\times (1 + O(g^2)).
\label{BB2}
\end{equation}
The corrections here are potentially large and 
are not suppressed by powers of $a$, so they cannot be
absorbed into the lattice corrections used in the continuum
extrapolation of mass ratios. So, wherever $\xi_r \equiv a_s/a_t$ 
is important it should be calculated explicitly.
This can be done straightforwardly (see Appendix D of
\cite{mt3d}
for example) if necessary. How necessary is it to do so? It is known
\cite{MPrenorm, trottier} 
that $\xi_r \simeq \xi$ for the type of improved
actions used here. So one will have an idea of the spatial 
volume that is sufficient for control of finite volume
effects, just from the tree-level estimate. Moreover one
does not need to know $\xi_r$ for ratios of glueball masses,
since the scale $a_t$ disappears in such ratios and the correction
terms will be affected in a minor way. However there
is at least one place where it is 
important and that is for calculations of how the potential
(which is obtained in units of $a_t$) varies with distance (which
is in units of $a_s$). So we need to know $\xi_r$ if
we wish to express the glueball masses in units of
$r_0$ or $\surd\sigma$. The quantity $r^2_0\sigma$,
on the other hand, is a dimensionless ratio of quantities
derived from the potential and so the precise value of
$\xi_r$ should be less important for it. Since  $\xi_r$
is not explicitly calculated in 
\cite{MPasym}
this means that we have to be cautious about quantities
that combine glueball masses with $r_0$ or $\sigma$. We
shall see below that such caution is indeed justified.

Now to the second question. The tadpole-improvement 
is supposed to ensure that the leading correction
is $O(\alpha_s a^2_s)$ rather than $O(a^2_s)$. For 
calculations where $a_s$ is large, as it is here,
one might argue that the  notionally sub-leading $O(a^4_s)$
correction will be much more important than the
notionally leading $O(\alpha_s a^2_s)$ correction, so 
that the latter can be neglected and the continuum
extrapolation can be performed using just the former.
This has an obvious danger however: it will produce very
small statistical errors at the price of potentially large
systematic errors. 

There is no \`a priori obvious answer to this second question 
so we turn to what the calculations themselves tell us.
The most accurately calculated mass is the lightest
$0^{++}$ and so we can ask what fit it prefers.
We consider the dimensionless quantity  $r_0 m_{0^{++}}$
and try to perform a continuum extrapolation with
corrections that are powers of $a_s/r_0$.
For the $\xi=3$ calculations we find that we can get 
a very good fit for $\beta\geq 2.0$, 
\begin{equation}
r_0 m_{0^{++}} = 4.10(20) 
- (5.0\pm1.2)\biggl({{a_s}\over{r_0}}\biggr)^2
+ (6.5\pm1.6)\biggl({{a_s}\over{r_0}}\biggr)^4,
\label{BB3}
\end{equation}
and poorer but still acceptable fits even if we include
all the values in Table~\ref{table_masym3_su3}.
For  $\xi=5$ we can fit the values in  
Table~\ref{table_masym5_su3} equally well:
\begin{equation}
r_0 m_{0^{++}} = 3.83(15) 
- (3.45\pm0.75)\biggl({{a_s}\over{r_0}}\biggr)^2
+ (4.50\pm0.80)\biggl({{a_s}\over{r_0}}\biggr)^4
\label{BB4}
\end{equation}
In neither case are statistically acceptable fits 
with just a $O(a^2_s)$ or just a $O(a^4_s)$ correction
possible, as should be apparent from eqns(\ref{BB3},\ref{BB4}). 
Note also that the coefficient of the $O(a^2_s)$ term is
not much smaller than that of the $O(a^4_s)$ term despite
the fact that it presumably comes with an
overall factor of $\alpha_s$. This is presumably because,
for these very coarse lattice spacings, $\alpha_s$ is not
small.

Although this already indicates that any continuum
fit needs to incorporate both $O(a^2_s)$ and $O(a^4_s)$
correction terms, there is the possibility that the lightest 
$0^{++}$ glueball is special. For example, its
mass vanishes at the nearby critical point in the
fundamental-adjoint coupling plane. (Although it is not
apparent why this should lead to a larg $O(a^2_s)$ correction
of the kind that we find.) Let us then
consider the continuum extrapolation of $r_0\surd\sigma$.
For $\xi=3$ we obtain a good fit, for $\beta \geq 1.9$,
as follows:
\begin{equation}
r_0 \surd\sigma = 1.192(10) 
- (0.32\pm0.08)\biggl({{a_s}\over{r_0}}\biggr)^2
+ (0.66\pm0.11)\biggl({{a_s}\over{r_0}}\biggr)^4
\label{BB5}
\end{equation}
As we shall see below, this continuum limit is
consistent with what one obtains from
the standard $a_s=a_t$ calculations. Once again 
fits with just a $O(a^2_s)$ or a $O(a^4_s)$ correction
term are not possible. 

We infer from the above analysis that continuum
extrapolations from this region of lattice spacings
should include both $O(a^2_s)$ and $O(a^4_s)$ corrections.
We can see from eqns(\ref{BB3}-\ref{BB5}) that
the coefficient of the $a^2_s$ term is somewhat smaller 
than that of the $a^4_s$ term, but that this difference is
not enough to make the $a^4_s$ correction the dominant one.
Thus we shall extrapolate all mass ratios using two
correction terms as in eqn(\ref{BB5}).

In the above discussion we ignored the fact
that we do not actually know the true value of $a_s/a_t$.
For the $r_0\surd\sigma$ extrapolation that does not
matter much, but for $r_0 m$ it does. The fact that it does
is apparent from the fits in eqns(\ref{BB3},\ref{BB4}).
If we take those values and translate to the string
tension using eqn(\ref{BB5}) we find 
\begin{equation}
{{m_{0^{++}}} \over {\surd\sigma}} =
\left\{ \begin{array}{ll}
3.44(17) & \ \ \ \xi=3 \\
3.21(13) & \ \ \ \xi=5
\end{array}
\right.
\label{BB6}
\end{equation}
Comparing with the value of $3.64\pm0.09$ 
in Table~\ref{table_mk_su2su3}, it is clear that
there is a problem with the values in eqn(\ref{BB6}).
The discrepancies are only $\sim5-10\%$ but this
is large at the level of statistical accuracy being
achieved in these calculations.

In conclusion, to minimise the systematic errors associated
with the continuum extrapolation, we shall use two correction
terms, as in  eqn(\ref{BB5}), in all our continuum extrapolations. 
To minimise the error associated with deviations of 
$\xi_r=a_s/a_t$ from the tree-level value, $\xi$, we shall
extrapolate to the continuum limit either ratios of glueball masses,
$a_tm_G/a_tm_{0^{++}} =  m_G/m_{0^{++}}$, or dimensionless
ratios of quantities from the potential, such as $r_0\surd\sigma$. 
To obtain the continuum value of $m_G/\surd\sigma$ we multiply
the extracted value of $m_G/m_{0^{++}}$ by the value of 
$m_{0^{++}}/\surd\sigma$ as obtained in the lattice calculations
with $a_s=a_t$.
If we do so we obtain the masses listed in Table~\ref{table_mcont_su3}.
(The $a_s\not=a_t$ values are obtained by averaging the
$\xi=3,5$ results, and for the $2^{++}$ we use both lattice
representations.)
Averaging these masses leads to the final mass estimates in
Table~\ref{table_mk_su2su3} and Table~\ref{table_mm_su2su3}.

\begin{table}[tbh]
\begin{center}
\begin{tabular}{|c||c|c||c|c|}\hline
\multicolumn{5}{|c|}{SU(3)} \\ \hline
\multicolumn{1}{|c||}{} &
\multicolumn{2}{c||}{$m_G/m_{0^{++}}$} & 
\multicolumn{2}{c|}{$m_G/\surd\sigma$} \\ \hline
 $G$  & $a_s\not=a_t$  & $a_s=a_t$  & $a_s\not=a_t$  & $a_s=a_t$
 \\ \hline
 $0^{++}$  &          &          &          & 3.64(9)   \\
 $2^{++}$  & 1.43(8)  & 1.40(9)  & 5.21(32) & 5.16(20)  \\
 $0^{-+}$  &          & 1.34(18) &          & 4.97(57)  \\
 $2^{-+}$  &          & 1.94(20) &          & 7.05(65)  \\
 $1^{+-}$  & 1.70(10) & 1.81(17) & 6.19(39) & 6.56(52)  \\
 $0^{++\star}$  &  1.78(14) & [1.77(21)] & 6.48(54) & [6.65(47)] \\
 $2^{++\star}$  &  1.85(20) &          & 6.73(73) &   \\ \hline
\end{tabular}
\caption{\label{table_mcont_su3}
Continuum mass ratios for the two different kinds of
calculation. Brackets denote continuum extrapolations
from only two values of $\beta$.}
\end{center}
\end{table}

To gain the full benefit of this kind of calculation, 
one clearly needs to calculate $\xi_r \equiv a_s/a_t$
at every value of $\beta$. This will have its own statistical
error and is likely to increase the final errors on the
masses by a significant amount. In addition
it would be useful to have calculations down to
smaller values of the lattice spacing (perhaps down
to the equivalent of $\beta=5.9$ or $\beta=6.0$ in the
isotropic action calculations) so as to achieve a much 
better control of the $c_1 a^2 + c_2 a^4$ extrapolation. 
A calculation possessing these features would produce a 
very accurate mass spectrum.

\vspace{0.35cm}

\noindent {\large {\bf (b) the $\bf r_0$ parameter}}

\vspace{0.35cm}

In addition to the calculations of $r_0/a$ from
\cite{MPasym}
which were listed in Table~\ref{table_masym3_su3}
and Table~\ref{table_masym5_su3} there have been
two recent calculations using the Wilson
plaquette action
\cite{ehkr0,alphar0}
and also a scattering of older calculations, including
\cite{bs1,bs2,ukqcd3,sesam,ukqcd4}. 
These are listed in Table~\ref{table_r0_su3}.

\begin{table}[tbh]
\begin{center}
\begin{tabular}{|c|l|l|l|}\hline
\multicolumn{4}{|c|}{SU(3) ;  $ \ \ r_0/a$} \\ \hline
 $\beta$  & ref\cite{alphar0}  &  ref\cite{ehkr0} & 
ref\cite{sesam,bs1,bs2,ukqcd3,ukqcd4} \\ \hline
 5.54 &           & 2.054(13) &  \\
 5.60 &           & 2.344(8)  &  \\
 5.70 & 2.922(9)  & 2.990(24) &  \\
 5.80 & 3.673(5)  &           &  \\
 5.85 &           & 4.103(12) &  \\
 5.95 & 4.898(12) &           &  \\
 6.00 &           & 5.369(9)  & $5.35(3)$   \\
 6.07 & 6.033(17) &           &  \\
 6.20 & 7.380(26) &           & $7.37(3)$   \\
 6.40 & 9.74(5)   &           & $9.89(16)$  \\
 6.40 &           &           & $9.75(17)$  \\
 6.50 &           &           & $11.23(21)$  \\
 6.57 & 12.38(7)  &           &  \\ \hline
\end{tabular}
\caption{\label{table_r0_su3}
Values of $r_0/a$ from calculations with the Wilson
plaquette action.}
\end{center}
\end{table}

We now separately extrapolate to the continuum limit the lattice 
values in each of the three columns of Table~\ref{table_r0_su3},
using an $a^2/r^2_0$ correction. (One gets almost identical
fits using an $a^2\sigma$ correction.) The continuum limits
are listed in Table~\ref{table_r0cont_su3}. We also show 
the continuum limits for the anistropic calculations listed
in  Tables~\ref{table_masym3_su3},~\ref{table_masym5_su3}.
In this case we use a $c_1 (a/r_0)^2 + c_2 (a/r_0)^4$
correction since, as we remarked earlier, we cannot
get statistically acceptable fits with just an $a^2$ or
an $a^4$ term. The $\xi=5$ calculation involves a fit with
3 parameters to 3 points and so we cannot tell how good a fit 
it is. I have therefore placed it in brackets. In the last column
I show the range of $\beta$ fitted in each case, as well
as the confidence level of the best fit and, in brackets, 
the CL corresponding to the error.

We note that all the fits are very good apart from that using 
the calculations of
\cite{ehkr0}.
One might imagine that this is because this calculation
goes down to lower values of $\beta$ where the $O(a^2)$
correction becomes insufficient. However if we just fit
the $\beta\geq 5.70$ part of the data we get a value
of 1.179(13) with an even worse confidence level, $\sim 20\%$.
The fact that the fit is worse closer to the continuum limit
is worrying and we therefore do not give this fit as
much weight as the three other fits. From these it seems
that a final value of 1.195(10) is a safe one and so this
is what we have quoted in Table~\ref{table_rest_su2su3}.

\begin{table}[tbh]
\begin{center}
\begin{tabular}{|c|l|l|c|}\hline
\multicolumn{4}{|c|}{SU(3)} \\ \hline
   &  $\lim_{a\to 0} r_0\surd\sigma$  & CL\% & range  \\ \hline
 \cite{alphar0} &  1.197(11) & 70(20) &
$5.70\leq\beta\leq 6.57$ \\
 \cite{ehkr0}   &  1.175(7)  & 40(10) &  
$5.54\leq\beta\leq 6.00$ \\
 \cite{bs1,bs2,ukqcd3,sesam,ukqcd4} & 1.204(26) & 95(25) & 
$6.0 \leq\beta\leq 6.5$ \\
 \cite{MPasym} $\xi=3$ &  1.192(10) & 75(25) &
$1.9 \leq\beta\leq 2.6$ \\
 \cite{MPasym} $\xi=5$ &  [1.177(13)] & --($\chi^2=1$) &
$1.9 \leq\beta\leq 2.4$ \\ \hline
\end{tabular}
\caption{\label{table_r0cont_su3}
Continuum values of $r_0\surd\sigma$ from various lattice
calculations.}
\end{center}
\end{table}

\vspace{0.35cm}

\noindent {\large {\bf (c) the deconfining temperature {$\bf T_c$}}}

\vspace{0.35cm}

As described in Appendix A, a periodic Euclidean lattice 
of size $N^3_s N_\tau$
with $N_s \gg N_\tau$ may be used to calculate the
properties of the gauge theory at a 
temperature  $T=1/aN_\tau$. Thus the deconfining temperature
$T_c$ may be obtained at fixed $N_\tau$ by varying $\beta$
and finding the value $\beta_c(N_\tau)$ at which the
system undergoes the deconfining phase transition. The 
corresponding temperature will be $a(\beta_c)T_c = 1/N_\tau$ 
in lattice units. There now exist calculations of this quantity
using a variety of actions.

The most extensive calculations are for the standard
plaquette action and these are listed in Table~\ref{table_Tc_su3}. 
As well as the critical couplings, $\beta_c$, I list 
my interpolated values of the string tension. 
The $N_\tau = 4,6$ calculations are from 
\cite{TcIsu3}
while the $N_\tau = 8,12$ calculations are from 
\cite{TcBsu3}.
The values I give are those obtained after an extrapolation to 
infinite volume using the finite-size scaling relation
\cite{TcIsu3,TcFsu3}
\begin{equation}
\beta_c(N_\tau,N_s)
=
\beta_c(N_\tau,\infty)
-
h \biggl({{N_\tau}\over{N_s}}\biggr)^3.
\label{BB7}
\end{equation}
For $N_\tau=4,6$ this relation 
has been fitted to a range of spatial volumes and
the coefficient $h$ explicitly determined
\cite{TcIsu3,TcFsu3}.
For $N_\tau=8,12$ there is no such direct calculation
but a value $h=0.06\pm0.04$ seems safe and this 
is what I have used in going from the values calculated in
\cite{TcBsu3}
for $N_\sigma=32$, to the values listed in 
Table~\ref{table_Tc_su3}. (Aside: the values of the
string tension used in 
\cite{TcBsu3}
were, in retrospect, incorrect. This led to the
continuum fit quoted there having an
unacceptably low confidence level, and so it should be 
disregarded. This has also been noted in
\cite{TcBImpsu3}
and their alternative continuum extrapolation is
consistent with ours.) I have extrapolated 
these lattice values to the continuum limit using
a $(aT_c)^2$ correction. The result, with confidence
level, is given in Table~\ref{table_contTc_su3}

\begin{table}[tbh]
\begin{center}
\begin{tabular}{|c|c|c|}\hline
\multicolumn{3}{|c|}{SU(3)} \\ \hline
  $N_\tau$ & $\beta_c(N_\tau, V=\infty)$ & $a\surd\sigma$ \\ \hline
  4  & 5.6925(2)  &  0.395(4)   \\
  6  & 5.8941(5)  &  0.265(2)   \\
  8  & 6.0618(11) &  0.1975(18) \\
 12  & 6.3364(25) &  0.1325(13) \\ \hline
\end{tabular}
\caption{\label{table_Tc_su3}
Values of the deconfining critical coupling, $\beta_c$,
and the string tension.}
\end{center}
\end{table}

There are in addition four other recent calculations
that use different improved actions
\cite{TcBImpsu3,TcTarosu3,TcBlisssu3,TcIwasu3}.
I have extrapolated these separately to the continuum
limit and the results are presented in 
Table~\ref{table_contTc_su3}. I have used an $O(a^2)$
correction in all cases. If I use instead an $O(a^4)$ 
correction, then the extrapolations do not move much -- 
certainly not outside the errors quoted in 
Table~\ref{table_contTc_su3}. The main change is that
the errors themselves are  typically halved.

What we see from Table~\ref{table_contTc_su3} is
that there is one group of calculations that
favours a value $ T_c/\surd\sigma \simeq 0.630$
and another group that favours 
$T_c/\surd\sigma \simeq 0.655$. The
statistical errors are small enough for these
two values to be incompatible. It is not possible to
make a completely rational choice between these
two sets of calculations. However, given the much larger
range of $N_\tau$ for the calculations with the
plaquette action, and given the accuracy with
which the string tension is known in that case,
it is hard not view the set to which it belongs as
being the more likely to be right. Nonetheless
we wish to play safe, so we choose for
Table~\ref{table_rest_su2su3} the value
$T_c/\surd\sigma = 0.640\pm 0.015$ which
encompasses the various possibilities.

\begin{table}[tbh]
\begin{center}
\begin{tabular}{|c|l|c|l|}\hline
\multicolumn{4}{|c|}{SU(3)} \\ \hline
  refs & $\lim_{a\to0} T_c/\surd\sigma$ & CL\% & range \\ \hline
  \cite{TcFsu3,TcIsu3,TcBsu3} & 
0.629(6)  &  75(25) & $N_\tau = 4,6,8,12$  \\
  \cite{TcBImpsu3} & 
0.634(8)  &  55(15) & $N_\tau = 3,4,5,6$  \\ 
  \cite{TcTarosu3} & 
0.627(12) &  -($\chi^2=1$) & $N_\tau = 4,6$  \\ 
  \cite{TcBlisssu3} & 
0.659(8)  &  85(25) & $N_\tau = 2,3,4$  \\ 
  \cite{TcIwasu3} & 
0.651(12) &  45(15) & $N_\tau = 3,4,6$  \\ \hline 
\end{tabular}
\caption{\label{table_contTc_su3}
Continuum value of the deconfining temperature in
units of the string tension, from the calculations
as shown.}
\end{center}
\end{table}

\vspace{0.35cm}

\noindent {\large {\bf (d) the topological susceptibility {$\bf \chi_t$}}}

\vspace{0.35cm}

As for SU(2) there are a number of quite different ways
of calculating the topological charge, $Q$, and the resulting
topological susceptibility, 
$\chi_t \equiv \langle Q^2 \rangle /V$. The older calculations
\cite{myQoldsu3}
use cooling; after typically 20 cooling sweeps they calculate
the lattice topological charge density, $Q_L(x)$, whose sum over $x$
provides an estimate for $Q$. Once again we can either take the 
full susceptibility, $\chi_c$, or we can try to remove the narrow 
instantons that might be lattice artifacts. As in the case of SU(2) 
we shall present results where all instantons with 
$Q_L(x_{peak}) \geq 1/16\pi^2$ have been removed; the corresponding
susceptibility we call $\chi_{c^\prime}$. (As for SU(2) we suppress
the subscript $t$ on $\chi$ for clarity.) There is also a
recent calculation that uses cooling 
\cite{myQnewsu3}.
As for SU(2) there are calculations that work directly
with the Monte Carlo generated gauge fields, but use
smeared operators and correct explicitly for the additive
and multiplicative renormalisations
\cite{pisaQsu3}.
We call the resulting susceptibility $\chi_{sm}$.
There are also calculations
\cite{boulderQsu3}
that use a smoothing of the gauge fields that is motivated by 
renormalisation  group considerations. This gives $\chi_{RG}$. 
At $\beta=6.0$ two of the above calculations provide values on 
two lattice sizes. The values of $a\chi^{1/4}_t$ corresponding 
to all these calculations are listed in Table~\ref{table_Q_su3}.

\begin{table}[tbh]
\begin{center}
\begin{tabular}{|c|c|c|c|c|c|}\hline
\multicolumn{6}{|c|}{SU(3)} \\ \hline
$\beta$ &  $a\chi^{1/4}_c$ \cite{myQoldsu3} &
 $a\chi^{1/4}_{c^\prime}$ \cite{myQoldsu3} &
 $a\chi^{1/4}_c$ \cite{myQnewsu3} &
 $a\chi^{1/4}_{sm}$ \cite{pisaQsu3}  & 
 $a\chi^{1/4}_{RG}$ \cite{boulderQsu3}  \\ \hline
5.70 & 0.1595(23) & 0.1244(33) &  &            &  \\
5.80 & 0.1347(25) & 0.1147(28) &  &            &  \\
5.85 &            &            &            &            & 0.1179(28) \\
5.90 & 0.1135(17) & 0.1059(16) &            & 0.1019(21) &  \\
6.00 & 0.0915(21) & 0.0880(22) & 0.0977(34) & 0.0901(19) & 0.0937(12) \\
6.00 &            &            & 0.0870(40) &            & 0.0961(33) \\
6.10 &            &            &            & 0.0799(17) & 0.0816(20) \\
6.20 & 0.0722(36) & 0.0713(35) & 0.0701(25) &            &  \\
6.40 &            &            & 0.0537(48) &            &  \\ \hline
\end{tabular}
\caption{\label{table_Q_su3}
The topological susceptibility, using various methods as defined
in the text.}
\end{center}
\end{table}

Extrapolating to the continuum limit using an 
$O(a^2\sigma)$ correction I obtain the continuum values
shown in Table~\ref{table_Qcont_su3}. These are clearly
consistent with each other. In addition, apart from the first
two columns which come from the same configurations, these
calculations are statistically independent (apart from the
string tension). This provides the final continuum estimate in
Table~\ref{table_rest_su2su3}.

\begin{table}[tbh]
\begin{center}
\begin{tabular}{|c|c|c|c|c|}\hline
\multicolumn{5}{|c|}{SU(3) : $\lim_{a\to 0}\chi_n/\surd\sigma$} \\ \hline
$n=c$ \cite{myQoldsu3}  & $n=c^\prime$ \cite{myQoldsu3} 
& $n=c$ \cite{myQnewsu3} &  $n=sm$  \cite{pisaQsu3} 
& $n=RG$  \cite{boulderQsu3} \\ \hline
 0.439(16) & 0.459(17) & 0.448(50) & 0.464(23) & 0.456(23) \\ \hline
\end{tabular}
\caption{\label{table_Qcont_su3}
The continuum topological susceptibility, extrapolating the values in
Table~\ref{table_Q_su3}, in units of the string  tension.}
\end{center}
\end{table}

\vspace{0.35cm}

\noindent {\large {\bf (e) the {$\bf \Lambda_{\overline{MS}}$} scale}}

\vspace{0.35cm}

To calculate $\Lambda_{\overline{MS}}$ one needs an accurate 
calculation of the running coupling over a large range of
length scales. Some of the first realistic attempts involved 
calculations of the static potential, with the running coupling 
being obtained from the Coulomb interaction at short distances. 
(See for example
\cite{ukqcdlambda}.)
The most recent calculation uses a different technique that
allows a fine control of the various systematic errors
\cite{alphalambda}
and this is the value we have quoted in 
Table~\ref{table_rest_su2su3}.

\vspace{1.0cm}


\begin{thebibliography}{99}

\bibitem{dalley}
S. Dalley and B. van de Sande, 
hep-th/9810236, hep-th/9806231.

\bibitem{AdS}
C. Csaki, H. Ooguri, Y. Oz and J. Terning,
hep-th/9806021. \\
H. Ooguri, H. Robins and J. Tannenhauser, 
hep-th/9806171.

\bibitem{Malda}
J. Maldacena, hep-th/9711200; hep-th/9803002. \\
S-J Rey and J. Yee, hep-th/9803001. \\
E. Witten, hep-th/9803131; hep-ph/9812208.

\bibitem{mt3d}
M. Teper, hep-lat/9804008 to appear in Phys. Rev. D.

\bibitem{Tc3d}
M. Teper, Phys. Lett. B313 (1993) 417 and unpublished. \\
J. Engels et al, Nucl. Phys. Proc. Suppl. 53 (1997) 420. \\
C. Legeland et al, Nucl. Phys. Proc. Suppl. 63 (1998) 260. 

\bibitem{symanzik}
K. Symanzik, Nucl. Phys. B226 (1983) 187; 205.

\bibitem{ukqcd64}
G. Bali et al (UKQCD Collaboration), 
Phys. Lett. B309 (1993) 378.

\bibitem{MPnew}
M. Peardon, hep-lat/9710029. \\
C. Morningstar and M. Peardon, hep-lat/9808045.

\bibitem{ukqcd285}
S. Booth et al (UKQCD Collaboration), 
Nucl. Phys. B394 (1993) 509.

\bibitem{MTnewton}
M. Teper,
Newton Inst. NATO-ASI School Lectures, July 1997, hep-lat/9711011.

\bibitem{mtsu4}
M. Teper, Phys. Lett. B397 (1997) 223;  
Nucl. Phys. Proc. Suppl. 53 (1997) 715,  and unpublished.

\bibitem{suN}
G. 't Hooft, Nucl. Phys. B72 (1974) 461. \\
E. Witten, Nucl. Phys. B160 (1979) 57. \\
S. Coleman, 1979 Erice Lectures. \\
A. Manohar, 1997 Les Houches Lectures, hep-ph/9802419.

\bibitem{parisi}
G. Parisi, in {\it High Energy Physics} - 1980(AIP 1981).

\bibitem{lepage}
P. Lepage and P. Mackenzie, Phys. Rev. D48 (1993) 2250. \\
P. Lepage, 1996 Schladming Lectures: hep-lat/9607076. 

\bibitem{ip}
N. Isgur and J. Paton, Phys. Rev. D31 (1985) 2910. \\
T. Moretto and M. Teper, hep-lat/9312035. \\
R. Johnson and M. Teper, hep-lat/9709083.

\bibitem{allJ}
R. Johnson and M. Teper, hep-lat/9808012.

\bibitem{myasat}
K. Ishikawa, G. Schierholz and M. Teper,
Z. Phys. C19 (1983) 327.

\bibitem{wilson}
K. Wilson, Phys. Rev. D10 (1974) 2445.

\bibitem{mtunpub}
M. Teper, unpublished.

\bibitem{cmmtsu2}
C. Michael and M. Teper, Nucl. Phys. B305 (1988) 453.

\bibitem{cmsu2}
C. Michael and S. Perantonis, J. Phys. G18 (1992) 1725.

\bibitem{ukqcdsu2}
S. Booth et al (UKQCD), Nucl. Phys. B394 (1993) 509.

\bibitem{balisu2}
G. Bali, K. Schilling, C. Schlichter and A. Wachter, 
hep-lat/9409005.

\bibitem{stringcor}
M. Luscher, K. Symanzik and P. Weisz,
Nucl. Phys. B173 (1980) 365. \\
Ph. de Forcrand, G. Schierholz, H. Schneider and M. Teper,
Phys. Lett. B160 (1985) 137. 

\bibitem{trottier}
N. Shakespeare and H. Trottier, hep-lat/9803024.

\bibitem{manton}
N. Manton, Phys. Lett. B96 (1980) 328.

\bibitem{sommersu2}
R. Sommer, Nucl. Phys. B411 (1994) 839.

\bibitem{sommerpc}
R. Sommer, private communication.

\bibitem{Tcsu2}
J. Fingberg, U. Heller and F. Karsch, 
Nucl. Phys. B392 (1993) 493 [hep-lat/9208012].

\bibitem{MTcool}
M. Teper, Phys. Lett. 162B (1985) 357.

\bibitem{cmmtQ}
C. Michael and M. Teper, Phys. Lett. 199B (1987) 95.

\bibitem{mtdp}
D. Pugh and M. Teper, Phys. Lett. 218B (1989) 326.

\bibitem{Zurichsu2}
Ph. de Forcrand, M. Garcia Perez and I-O Stamatescu,
hep-lat/9701012.

\bibitem{Bouldersu2}
T. DeGrand, A. Hasenfratz and T. Kovacs, hep-lat/9711032.

\bibitem{Pisasu2}
B. Alles, M. D'Elia and A. Di Giacomo, hep-lat/9706016.

\bibitem{fsstK}
Ph. de Forcrand, G. Schierholz, H. Schneider and M. Teper,
Phys. Lett. 160B (1985) 137.

\bibitem{ehkr0}
R. Edwards, U. Heller and T. Klassen,
hep-lat/9711003.

\bibitem{cmmtsu3}
C. Michael and M. Teper,
Nucl. Phys. B314 (1989) 347; and unpublished.

\bibitem{spcmsu3}
S. Perantonis and C. Michael,
Nucl. Phys. B347 (1990) 854.

\bibitem{bs1}
G. Bali and K. Schilling,
Phys. Rev. D46 (1992) 2636.

\bibitem{bs2}
G. Bali and K. Schilling,
Phys. Rev. D47 (1993) 661.

\bibitem{ukqcd1}
C. Allton et al (UKQCD),
Nucl. Phys. B407 (1993) 331.

\bibitem{ukqcd2}
S. Booth et al (UKQCD),
Phys. Lett. B294 (1992) 385.

\bibitem{ukqcd3}
H. Wittig (UKQCD),
Nucl. Phys. Proc. Suppl. 42 (1995) 288.

\bibitem{fsstM}
Ph. de Forcrand, G. Schierholz, H. Schneider and M. Teper,
Phys. Lett. 152B (1985) 107.

\bibitem{GF11}
H. Chen, J. Sexton, A. Vaccarino and D. Weingarten,
Nucl. Phys. Proc. Suppl. 34 (1994) 357.

\bibitem{MPasym}
C. Morningstar and M. Peardon, hep-lat/9704011.

\bibitem{MPrenorm}
C. Morningstar and M. Peardon, 
Nucl. Phys. Proc. Suppl. 53 (1996) 914.

\bibitem{alphar0}
M. Guagnelli, R. Sommer and H. Wittig (ALPHA),
hep-lat/9806005.

\bibitem{sesam}
U. Gl\"assner et al (SESAM),
Phys. Lett. B383 (1996) 98.

\bibitem{ukqcd4}
H. Wittig (UKQCD),
Int. J. Mod. Phys. A12 (1997) 4477.

\bibitem{TcIsu3}
Y. Iwasaki et al, 
Phys. Rev. D46 (1992) 4657.

\bibitem{TcBsu3}
G. Boyd et al,
Nucl. Phys. B469 (1996) 419 [hep-lat/9602007].

\bibitem{TcFsu3}
M. Fukugita, M. Okawa and A. Ukawa,
Nucl. Phys. B337 (1990) 181.

\bibitem{TcBImpsu3}
B. Beinlich, F. Karsch, E. Laermann and A, Peikart,
hep-lat/9707023.

\bibitem{TcTarosu3}
Ph. de Forcrand et al (QCD-TARO),
hep-lat/9809086.

\bibitem{TcBlisssu3}
D. Bliss, K. Hornbostel and G. Lepage,
hep-lat/9605041.

\bibitem{TcIwasu3}
Y. Iwasaki, K. Kanaya, T. Kaneko and T. Yoshie,
Phys. Rev. D56 (1997) 151.

\bibitem{myQoldsu3}
M. Teper Phys. Lett. 202B (1988) 553. \\
J. Hoek, M. Teper and J. Waterhouse,
Nucl. Phys. B288 (1987) 589.

\bibitem{myQnewsu3}
D. Smith and M. Teper,
Phys. Rev. D58 (1998) 014505.

\bibitem{pisaQsu3}
B. Alles, M. D'Elia and A. Di Giacomo,
Nucl. Phys. B494 (1997) 281.

\bibitem{boulderQsu3}
A. Hasenfratz and C. Nieter,
hep-lat/9806026.


\bibitem{ukqcdlambda}
D. Henty et al (UKQCD), 
Phys. lett. B294 (1992) 385.

\bibitem{alphalambda}
S. Capitani et al (ALPHA), hep-lat/9709125.

\end{thebibliography}
\end{document}